\documentclass[aps,prl,twocolumn,showpacs]{revtex4-1}
\usepackage{epsfig}
\usepackage{graphicx}
\usepackage{bm}
\usepackage{amssymb}
\usepackage{amsmath}
\usepackage{amsthm}
\begin{document}
\title{The effect of disorder on transverse domain wall dynamics in 
magnetic nanostrips}
\author{Ben Van de Wiele$^{1}$, Lasse Laurson$^2$, and Gianfranco Durin$^{3,4}$}
\affiliation{$^1$Department of Electrical Energy, Systems and Automation, Ghent 
University, Ghent B-9000, Belgium}
\affiliation{$^2$COMP Centre of Excellence, Department of Applied Physics, 
Aalto University, PO Box 14100, Aalto 00076, Finland}
\affiliation{$^3$ISI Foundation, Via Alassio 11/c, 10126 Torino, Italy}
\affiliation{$^4$Istituto Nazionale di Ricerca Metrologica, Strada delle Cacce 91, 
10135 Torino, Italy}
\begin{abstract}
We study the effect of disorder on the dynamics of a transverse domain wall 
in ferromagnetic nanostrips, driven either by magnetic fields or 
spin-polarized currents, by performing a large ensemble of GPU-accelerated 
micromagnetic simulations. Disorder is modeled by including small, randomly 
distributed non-magnetic voids in the system. Studying the 
domain wall velocity as a function of the applied field and current density
reveals fundamental differences in the domain wall dynamics induced by these 
two modes of driving: For the field-driven case, we identify two different 
domain wall pinning mechanisms, operating below and above the Walker breakdown, 
respectively, whereas for the current-driven case pinning is absent above 
the Walker breakdown. Increasing the disorder strength induces a larger Walker 
breakdown field and current, and leads to decreased and increased domain wall
velocities at the breakdown field and current, respectively. Furthermore, 
for adiabatic spin transfer torque, the intrinsic pinning mechanism is 
found to be suppressed by disorder. We explain these findings within the
one-dimensional model in terms of an effective damping parameter $\alpha^*$ 
increasing with the disorder strength.
\end{abstract}
\maketitle

Domain wall (DW) dynamics in nanoscale ferromagnetic wires and strips 
driven by magnetic fields or spin-polarized currents is a subject of major 
technological importance for the operation of potential future 
nanoscale magnetic memory \cite{PAR-08,HAY-08} and logic \cite{ALL-05} 
devices. In these devices information is typically stored as 
magnetic domains along a nanostrip/wire and is processed by DW 
motion. For the reliable operation of such devices it is of fundamental 
importance to understand and control the effect of imperfections or 
disorder on the DW dynamics, necessarily present
in any realistic samples, e.g. in the form of thickness fluctuations and 
grain structure of the sample, or various impurities and defects in the 
material. At the same time, such systems constitute a low-dimensional 
limit of the general problem of driven elastic manifolds in a random 
potential \cite{LEC-09}.

While the crucial importance of disorder for the dynamics of 
higher-dimensional DWs is well established, resulting in 
phenomena such as the Barkhausen effect \cite{DUR-06}, majority of 
studies of DW motion in systems with nanostrip/wire geometry 
neglect disorder effects. This applies to both theoretical studies 
and interpretations of experimental results. Some exceptions include 
studies demonstrating enhanced DW propagation due to  
roughness of the edges of the strip \cite{NAK-03,MAR-12}.
Recently also the effect of spatially varying saturation magnetization 
$M_s$ on the dynamics of vortex walls was studied, resulting in 
an effective damping increasing with the disorder strength \cite{MIN-10}. 
Similar spatially distributed disorder has also been studied in a 
simplified, line-based model of a transverse DW \cite{LAU-10,LAU-11}. 
Experimental studies of DW dynamics in wires have revealed its stochastic 
nature in the case of short current pulses \cite{MEI-07}, and has 
been attributed to the presence of disorder in the samples, in combination 
with thermal effects. For longer current pulses, the resulting average 
DW velocities have been shown to be quite low \cite{KLA-05}, likely due
to pinning effects induced by structural disorder. Dynamical pinning
effects have also been observed in experiments of field-driven vortex
wall dynamics \cite{TAN-08,JIA-10}. However, despite 
of these advances, many details of the disorder effects on DW dynamics 
in nanostructures remain to be clarified.

In this Letter, we consider by micromagnetic simulations the effect of 
disorder on the field and current-driven dynamics of a transverse DW in 
a narrow and thin Permalloy strip. Disorder is modelled by including
randomly positioned small non-magnetic regions (voids) in the system.
Our results show that the field and current-driven DW dynamics
exhibit remarkable differences which are only revealed in the presence
of disorder. In particular, we identify two fundamentally different DW 
pinning mechanisms acting in a field-driven system, operating below 
and above the Walker breakdown field, respectively, with the latter mechanism
being absent in the current-driven case. Also the Walker breakdown itself 
is affected by the presence of disorder, such that it is shifted to larger
field and current values with increasing disorder strength. At the same time
the DW velocities at the breakdown field and current get smaller 
and larger, respectively. Furthermore, for adiabatic spin transfer 
torque, the intrinsic pinning mechanism is found to be suppressed by 
disorder. These findings emphasize the importance of 
understanding the interplay between disorder, the DW structure and 
the properties of the external driving force, and are shown to be
related to an effective damping parameter $\alpha^*$ increasing with
the disorder strength. 

We perform a large ensemble of micromagnetic simulations with the GPU-based
micromagnetic simulator MuMax \cite{VAN-11},
making it possible to obtain large statistics for averaging 
over the disorder realizations. To study the time evolution of the magnetization
${\bf M}({\bf r},t)$ with an amplitude $M_s$, we solve the 
Landau-Lifshitz (LL) equation with the spin-transfer torque 
terms \cite{ZHA-04},
\begin{eqnarray}
\label{eq:1}
\frac{\partial {\bf M}}{\partial t} & = & -\frac{\gamma}{1+\alpha^2} 
{\bf{M}}\times {\bf H}_{eff} \\ \nonumber
& & -\frac{\alpha \gamma}{M_s(1+\alpha^2)} 
{\bf M}\times({\bf M}\times{\bf H}_{eff}) \\ \nonumber
& & -\frac{b_j}{M_s^2(1+\alpha^2)}{\bf M}\times ({\bf M}\times 
({\bf j}\cdot \nabla){\bf{M}}) \\ \nonumber 
& & -\frac{b_j}{M_s(1+\alpha^2)}(\xi-\alpha){\bf M} \times 
({\bf j}\cdot \nabla){\bf M},
\end{eqnarray}
where ${\bf H}_{eff}$ is the effective magnetic field (with contributions 
from the external, exchange and demagnetization fields), 
$\gamma$ is the gyromagnetic ratio, $\alpha$ is the Gilbert damping constant, 
$\xi$ is the degree of non-adiabaticity, ${\bf j}$ is the current density, 
and $b_j=P\mu_B/(eM_s(1+\xi^2))$, with $P$ the polarization, $\mu_B$ the Bohr 
magneton and $e$ the electron charge.

\begin{figure}[t!]
\includegraphics[width=8cm,clip]{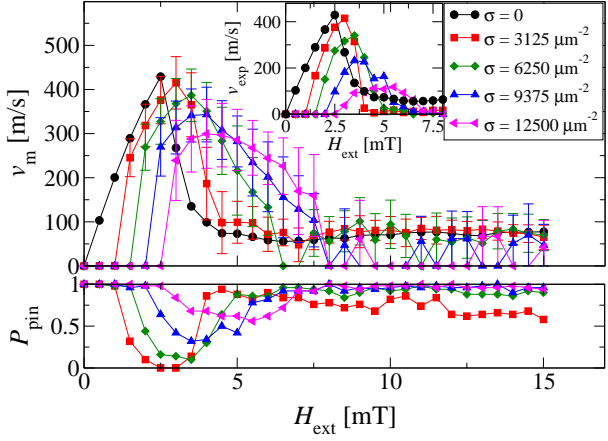}
\caption{(Color online) The average velocity $v_m$ of the moving DWs (main
figure) and $v_{exp} = (1-P_{pin})v_m$ (inset)
as a function of $H_{ext}$ and $\sigma$. Error bars correspond to the 
standard deviation of $v_m$. The pinning probabilities $P_{pin}$ 
during the 20 ns simulation (bottom panel) exhibit large values for 
large $H_{ext}$ due to the core pinning mechanism.}
\label{fig:field}
\end{figure}

We consider Permalloy strips of width $w=100$ nm 
and thickness $10$ nm, such that the stable DW structure is a 
head-to-head V-shaped symmetric transverse wall, separating in-plane 
domains pointing along the strip axis \cite{NAK-05}. The used material parameters are 
those of Permalloy, i.e. $M_s=860 \times 10^3$ A/m and $\alpha=0.02$,
and no anisotropy fields are included in Eq. (\ref{eq:1}).
To clearly see the effect of quenched disorder on the DW dynamics, we set 
the temperature $T=0$. The system is discretized by considering $N$ cells
of size $3.125 \times 3.125 \times 10$ nm$^3$. Upon application of an 
external magnetic field ${\bf H}_{ext} = H_{ext}\hat{\bf x}$ along the 
strip axis in the absence of disorder, the DW is displaced 
along the strip. If the field is below the Walker breakdown field $H_W$, 
the DW essentially keeps its equilibrium structure during the propagation, 
with a small out-of-plane component close to the tip of the V-shape, 
and a velocity roughly linearly proportional to the applied field. Above 
$H_W$, an antivortex is nucleated at the tip of the V-shape. It then 
propagates across the strip width, reversing the polarity of the DW 
magnetization. This process is repeated such that the DW polarity 
oscillates back and forth, dramatically decreasing the average 
DW velocity \cite{THI-06}.

With disorder included in the form of randomly positioned 
non-magnetic voids of linear size 3.125 nm with varying densities 
$\sigma$ 
within a strip of length $L$ = 3.2 $\mu$m, the DW can
get pinned even for non-zero applied fields \cite{remark1}. 
This makes measurement and even definition of the DW velocity a 
non-trivial task. Thus, in what follows we consider both the ``conditional 
velocities'' $v_m$ of the moving DWs, conditioned on the fact 
that the DWs will not get pinned during the time interval 
$\Delta t = 20$ ns we consider in the simulations (i.e. the DW will either
reach the end of the strip or it is still moving after $\Delta t = 20$
ns)\cite{v_m}, and the probability $P_{pin}$ for the DW to get pinned during
$\Delta t$.
These are computed by averaging over 50 disorder realizations for each
$H_{ext}$ and $\sigma$. Notice that here we consider a $T = 0$ system, such 
that a pinned DW cannot depin.
An alternative measure of the DW velocity (which is likely 
to be closer to typical experimental measurements where $T>0$) is given 
by $v_{exp} = (1-P_{pin})v_m$. In general, $P_{pin}$ will increase with the 
observation (time and length) scale, thus making also $v_{exp}$ a 
scale-dependent quantity.

\begin{figure}[t!]
\includegraphics[width=8cm,clip]
{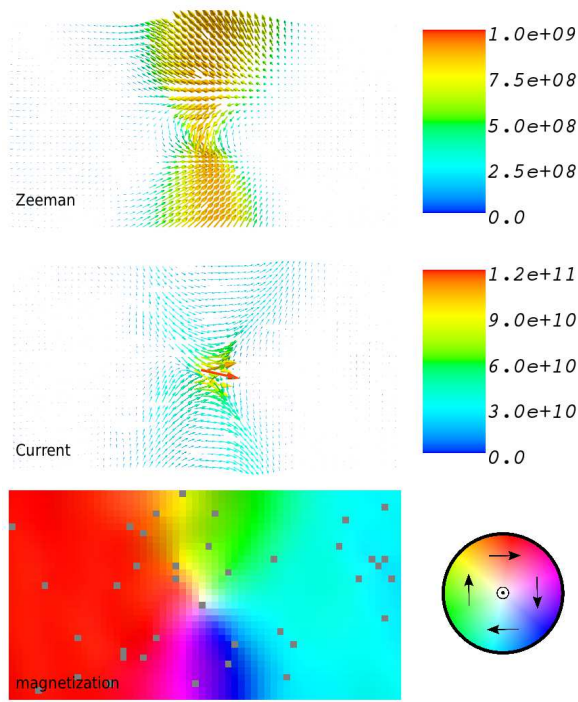}
\caption{(Color online) Examples of the spatial distribution
of the contributions of the applied field $H_{ext} =$ 5 mT (top) 
and current density $j_{ext} = 20 \times 10^{12}$ A/m$^2$ with 
$\xi=0$ (middle) to $\partial {\bf M}/\partial t$ 
in Eq. (\ref{eq:1}), corresponding to the magnetization 
configuration shown in the bottom panel, exhibiting an
antivortex in the middle of the strip. $\partial {\bf M}/\partial t$
is given in units of $M_s/$s. The randomly
positioned voids with  $\sigma = 3125 \mu$m$^{-2}$
are shown as grey dots.}
\label{fig:maps}
\end{figure}

Fig. \ref{fig:field} shows the resulting average velocities $v_{m}$
of the moving DWs as a function of $H_{ext}$ and $\sigma$. 
The presence of voids induces a finite depinning field 
$H_{dep}(\sigma)$ increasing with $\sigma$. For $H_{ext} > H_{dep}(\sigma)$, 
$v_m$ first increases until a maximum velocity is reached at 
$H_{ext} = H_W(\sigma)$, and then starts to decrease again. The position 
$H_W(\sigma)$ of this maximum, corresponding to the Walker breakdown, 
is shifted towards larger field values as $\sigma$ is increased, and 
the corresponding maximum velocity $v_{m}(H_W(\sigma))$ decreases with 
$\sigma$. The error bars in Fig. \ref{fig:field} correspond
to the standard deviation of $v_m$, and indicate
that the dynamics of moving DWs has a stochastic nature due 
to the random disorder. Notice in particular that the pinning probability
$P_{pin}$ exhibits a non-monotonic dependence on $H_{ext}$, with
strong pinning for both small and large $H_{ext}$, while for intermediate
applied fields (corresponding to large values of $v_m$) pinning is less 
likely. The maximum value of $v_{exp}$ (inset of Fig. \ref{fig:field})
exhibits a strong dependence on $\sigma$, and depends also on the
observation scale via $P_{pin}$ (not shown). For large $H_{ext}$, 
$P_{pin}$ is close to 1 for $\Delta t = 20$ ns, and consequently 
$v_{exp}$ is essentially zero. Similar pinning effects for large
applied fields have been observed experimentally for vortex walls
\cite{TAN-08,JIA-10}.

To gain insight on the mechanisms behind this behavior, we consider
snapshots of the DW configurations and the various contributions
to $\partial {\bf M}/\partial t$ in Eq. (\ref{eq:1}). 
For small $H_{ext}$, we find 
that the overall DW structure is preserved, with the disorder 
inducing only minor distortions. If the DW gets pinned, this happens 
by a collective action of several voids. This mechanism 
is known as {\it collective pinning}.
and it is responsible for the non-zero depinning field $H_{dep}<H_W(\sigma)$. 
Remarkably, we identify a fundamentally different pinning mechanism 
for large fields, $H_{ext} > H_W(\sigma)$: In this regime, an antivortex 
is able to propagate to the interior of the strip, 
resulting in pinned DW configurations (occurring with probability $P_{pin}$) 
with the antivortex core positioned exactly on top of a void or a local
void structure. We refer to this mechanism as {\it core pinning}, and 
attribute it to the fact that the energy of the system can be significantly 
lower when the antivortex core or part of it - involving large 
magnetization gradients and out-of-plane magnetization - is 
placed in a non-magnetic region (or more generally, in a region
with low $M_s$). In the field-driven case the DW
is susceptible to get pinned by this mechanism because the Zeeman 
torque is relatively small in magnitude and does not directly displace 
the DW (top panel of Fig. \ref{fig:maps}); Instead, the small 
out-of-plane magnetization due to the Zeeman torque induces demagnetizing 
fields, which act to move the DW. Such an indirect driving mechanism is 
sensitive to the perturbations due to disorder, leading to several effects, 
including $\sigma$-dependent $H_{dep}$ and $H_W$, and in particular the core
pinning mechanism for high $H_{ext}$. 

\begin{figure}[t!]
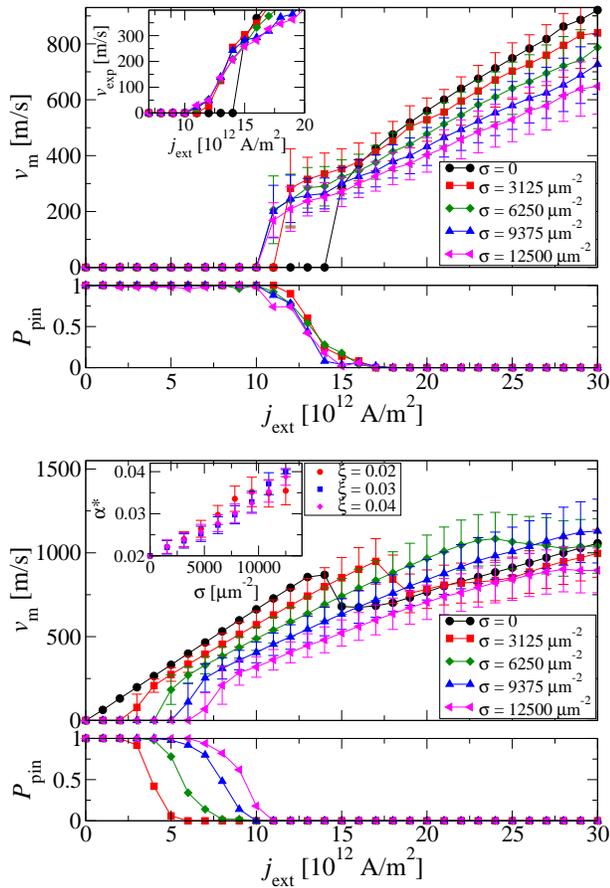

\includegraphics[width=8cm,clip]
{./velocity_xi=0.eps}\\
\vspace{0.3cm}
\includegraphics[width=8cm,clip]
{./velocity_xi=0.04_3.eps}
\caption{(Color online) The average velocity $v_m$ of 
the moving DWs as a function of $j_{ext}$ and $\sigma$,
for $\xi = 0$ (top) and $\xi = 0.04$ (bottom). 
Error bars correspond to the standard deviation of $v_m$.
The pinning probabilities $P_{pin}$ during the 20 ns simulation 
highlight the absence of core pinning for large current densities. The 
insets show $v_{exp} = (1-P_{pin})v_m$ for $\xi=0$ (top panel), and
the effective $\alpha^*(\sigma)$ for various $\xi$ (bottom panel), 
respectively.}
\label{fig:current}
\end{figure}

We proceed to contrast these results with the current-driven case, 
by applying a current density ${\bf j}=-j_{ext}\hat{\bf x}$ with $P=0.5$
along strips of length $L$=6.4 $\mu$m. We first consider perfect adiabaticity 
($\xi = 0$, top panel of Fig. \ref{fig:current}). Due to intrinsic 
pinning \cite{LI-04}, there is a non-zero depinning current $j_{dep,int}$
in the absence of disorder, above which DW motion 
involves repeated polarity transformations mediated by antivortex propagation 
across the strip width. Adding disorder with the same procedure as 
above reveals two intriguing observations: First, it appears that 
the DW is able to move even for currents slightly below 
$j_{dep,int}$. This surprising finding can be explained by noticing 
that the intrinsic pinning mechanism is due to the ability of the 
DW to deform in such a way that the torques due to interactions 
within the DW (i.e. the effective field) exactly counterbalance 
the adiabatic spin-transfer torque \cite{LI-04}. However, the presence of 
disorder induces additional DW deformations and imposes constraints on 
the ability of the DW to counteract the current-induced torques, 
leading to non-zero values for both $v_m$ and $1-P_{pin}$ for $j_{ext}$ 
somewhat below $j_{dep,int}$. Notice that while $v_{exp}$ (inset of the 
top panel in Fig. \ref{fig:current}) exhibits non-linear field dependence 
reminiscent of typical creep motion for small fields, we are considering 
here a $T = 0$ system in which a pinned DW cannot depin due to 
the absence of thermal fluctuations \cite{remark2}.

The second observation is that for larger $j_{ext}$, 
core pinning is absent. Even if for $j_{ext}>j_{W}(\sigma)$ the antivortex 
core is constantly moving back and forth across the strip width, it never 
gets pinned by the voids, strongly contrasting with the field-driven
case. To explain this observation, we consider the 
spatial distribution of the current-induced contribution to 
$\partial {\bf M}/\partial t$ (middle panel of Fig. \ref{fig:maps}), and 
find that the current acts directly (in contrast to the indirect mechanism
in the field-driven case) and strongly on the antivortex core 
where the magnetization gradients are large, facilitating its
propagation along the strip across the energy barriers 
due to the voids. This is also directly visible in the 
the LL equation (Eq. (\ref{eq:1})), where 
the current acts on the gradient of ${\bf M}$ rather than on ${\bf M}$ 
itself. 

Finally we consider the role of the non-adiabatic spin-transfer 
torque (bottom panel of Fig. \ref{fig:current}, where
the $\xi=0.04$ case is shown) on 
the DW dynamics. For $\xi>0$ and $\sigma=0$, there is no 
intrinsic pinning, and the DW propagates preserving its 
internal structure with a finite velocity linearly proportional to 
the current density $j_{ext}$ up to a Walker breakdown current $j_W$. 
For $j_{ext} > j_W$, an antivortex is again nucleated and propagates 
across the strip width reversing the polarity of the DW 
magnetization, and decreasing the average DW velocity. 
For larger $j_{ext}$, the velocity again increases with $j_{ext}$. 
Adding disorder induces a finite depinning threshold 
$j_{dep} (\sigma)$, and pushes the local maximum of $v_m$ or 
the Walker breakdown to higher $j_{ext}$. At the same time, 
$v_m$ at $j_{W}(\sigma)$ increases with $\sigma$. Thus, 
the voids are able to inhibit the antivortex entering the strip, 
enhancing the DW propagation and structural stability for intermediate 
current densities, $j_W(\sigma=0) < j_{ext} < j_W(\sigma>0)$. This 
effect arises as the antivortex core is pushed across the strip
width by the effective field terms in Eq. (\ref{eq:1}) (notice
that the effect of the current is symmetric such that no antivortex
displacement along the $y$ direction arises directly due to the
current, see the middle panel of Fig. \ref{fig:maps}), a mechanism
sensitive to the disturbances due to disorder. 
Again, there is no core pinning for $j_{ext} > j_W(\sigma)$, for the 
same reason as in the adiabatic ($\xi = 0$) case. 

For $j_{dep}(\sigma) < j_{ext} < j_W(\sigma)$, $v_m$  
depends linearly on $j_{ext}$, and by extrapolating linear 
fits to the data to $j_{ext}=0$ all the lines cross at $v_m = 0$ 
(not shown). Thus, we estimate effective values of the damping 
parameter from the slopes of these linear fits \cite{MIN-10}, as 
within one-dimensional models \cite{MOU-07} 
$v_m \propto (\beta/\alpha) j_{ext}$ for $j_{ext} < j_W$, with 
$\beta = \xi/(1+\xi^2)$. Our simulations (inset of the lower panel 
of Fig. \ref{fig:current}) with different $\xi$ indicate that the data 
can be interpreted in terms of an effective $\alpha^*$ increasing 
with $\sigma$ \cite{MIN-10}. Also an effective $M_s^* = (1-\sigma Lw/N)M_s$ 
emerges naturally. Thus we can explain our results with the one-dimensional 
model in terms of $\sigma$-dependent effective parameters: For instance, 
$j_W(\sigma) = 4\pi\gamma (M_s^2\Delta |N_y-N_x|)^*\alpha^*/(g\mu_BP|\beta-\alpha^*|)$,
with $\Delta$ the DW width and $N_x$ and $N_y$ the demagnetizing factors, 
and $j_{dep,int}(\sigma) \equiv j_W(\sigma,\xi=0)$
\cite{MOU-07}. Using the expression for $j_W$ and the values of $\alpha^*$ 
to estimate $C^*\equiv(\Delta M_s^2|N_y-N_x|)^*$, the 
scaling of $j_{dep,int}$ with $\sigma$ can be reproduced 
remarkably well, see Table \ref{table}. A similar analysis in the field-driven 
case, with $H_W = 2\pi\alpha^*(M_s|N_y-N_x|)^*$ and 
$v_m(H_W)=(\gamma \Delta^*/\alpha^*)H_W$ \cite{MOU-07}, 
reproduces the observed scaling of both $H_W$ and $v_m(H_W)$ 
with $\sigma$ (Table \ref{table}). Notice that in our case $v_m(H_W)$ 
depends on $\sigma$ through the $\sigma$-dependent effective parameters, 
while for systems with only edge roughness $v_m(H_W)$ is independent of the 
amount of edge roughness \cite{NAK-03}.  

\begin{table*}
\begin{center}
  \begin{tabular}{|c|c|c|c|c|c|c|c|c|}
    \hline
    $\sigma$ [$\mu$m$^{-2}$] & $\alpha^*$ & $C^*$ [A$^2$/m] & $j_{dep,int}^{pred}$ [A/m$^2$] & $j_{dep,int}^{sim}$ [A/m$^2$] & $H_W^{pred}$ [mT] & $H_W^{sim}$ [mT] & $v_m^{pred}(H_W)$ [m/s] & $v_m^{sim}(H_W)$ [m/s]\\ \hline
    0 & 0.0200 & 2.92 $\times 10^{-10}$ & 14 $\times 10^{12}$ & 15 $\times 10^{12}$ & 2.75 & 2.75 & 457 & 457\\ 
    1562.5 & 0.0221 & 2.52 $\times 10^{-10}$ & 12.1 $\times 10^{12}$ & 13 $\times 10^{12}$ & 3.05 & 3.0 & 398 & 437\\ 
    3125 & 0.0238 & 2.45 $\times 10^{-10}$ & 11.7 $\times 10^{12}$ & 12 $\times 10^{12}$ & 3.25 & 3.25 & 389 & 419\\ 
    4687.5 & 0.0258 & 2,36 $\times 10^{-10}$ & 11.3 $\times 10^{12}$ & 11.5 $\times 10^{12}$ & 3.52 & 3.25 & 377 & 403\\
    6250 & 0.0283 & 2.28 $\times 10^{-10}$ & 10.9 $\times 10^{12}$ & 11 $\times 10^{12}$ & 3.78 & 3.5 & 368 & 387\\ 
    \hline
  \end{tabular}
\end{center}
\caption{Predictions for $j_{dep,int}$, $H_W$ and $v_m(H_W)$ 
from the one-dimensional model in terms of $\sigma$-dependent 
effective $\alpha^*$ and $C^*\equiv(\Delta M_s^2|N_y-N_x|)^*$, 
compared with the simulated values. $C^*$ is estimated by fitting 
the expression for $j_W$ (see text) to the data in the bottom 
panel of Fig. \ref{fig:current}.}
\label{table}
\end{table*}

To summarize, we have presented a detailed analysis of the effect of 
disorder on the field and current-driven transverse DW 
dynamics in a narrow and thin Permalloy nanostrip. We have 
identified two fundamentally different pinning mechanisms, acting 
in different regimes of the DW propagation. The observation 
that there is no core pinning in the current-driven case whereas it 
dominates the field driven dynamics for large fields highlights the 
different nature of the field and current drive in a way that can 
be observed only in the presence of disorder. In general, we have
seen that the pinning mechanisms operating will depend on the 
details of the DW structure, and thus we expect that the 
core pinning mechanism is absent for systems with high perpendicular 
magnetocrystalline anisotropy as there is no (anti)vortex core that 
could get pinned, but it could play a role in the dynamics of vortex 
walls occurring in wider soft strips \cite{MIN-10}, possibly also 
for small applied fields. If only edge roughness is present, no
core pinning should occur. Experiments should be performed to 
systematically study the scale dependence of $P_{pin}$ and $v_{exp}$. 
Finally, 
we point out that the observation that disorder tends to stabilize 
the DW internal structure and increase the maximum DW velocity by 
suppressing the Walker breakdown in the current-driven case suggests 
that it could be desirable to deliberately engineer disorder in 
the system, for instance to replace notches to pin the DW in various 
technological applications \cite{BAS-12}.

{\bf Acknowledgments}. 
Stefano Zapperi is thanked for numerous interesting discussions on 
DW dynamics and disorder, and Mikko J. Alava for useful comments on
the manuscript. We thank Luc Dupr\'e and Dani\"el De Zutter for 
supporting this research. LL has been supported by the Academy of 
Finland through a Postdoctoral Researcher's Project (project no. 
139132) and through the Centres of Excellence Program (project no. 
251748). BVdW has been supported by the Flanders Research Foundation 
FWO.


\begin{thebibliography}{10}
\bibitem{v_m}
To avoid any effect related to the initial acceleration and of the
demagnetizing fields at the end of the wire, we actually calculated the average
speed between a point at 0.5 $\mu m$ after the initial position and at 0.5 $\mu
m$ before the end of the wire.
\bibitem{remark1}
We have checked that our main conclusions remain the same if instead
of voids one considers small areas with half the saturation magnetization
$M_s$, suggesting that our results are not limited to the specific 
kind of disorder we study here.
\bibitem{remark2}
In fact, studying the true thermally activated creep motion by micromagnetic
simulations is very challenging, due to the long time scales needed to
observe several repeated pinning-depinning events, a requirement for 
reliable estimation of the DW velocities in the creep regime.
\end{thebibliography}
\end{document}